\documentclass[aps,prb,twocolumn,amsmath,amssymb]{revtex4-1}

\usepackage{amsmath}
\usepackage{bm}
\usepackage{xcolor}
\usepackage{graphicx}

\usepackage[unicode=true,colorlinks=true]{hyperref}

\begin{document}

\title{Virtual crystal description of III-V semiconductor alloys in the tight binding approach}
\author{M.O.~Nestoklon}
\affiliation{Ioffe Physical-Technical Institute, 194021 St. Petersburg, Russia}
\affiliation{Laboratoire de Photonique et Nanostructures, CNRS and Universit{\'e} Paris-Saclay, Route de Nozay, 91460 Marcoussis, France}
\author{R.~Benchamekh}
\affiliation{Tyndall National Institute, Lee Maltings, Dyke Parade, Cork, Ireland}
\author{P.~Voisin}
\affiliation{Laboratoire de Photonique et Nanostructures, CNRS and Universit{\'e} Paris-Saclay, Route de Nozay, 91460 Marcoussis, France}

\begin{abstract}
We propose a simple and effective approach to construct the 
empirical tight-binding parameters of ternary alloys in the virtual crystal approximation. This combines
a new, compact formulation of the strain parameters
and a linear interpolation of the hamiltonians of binary materials \textit{strained to the alloy equilibrium lattice parameter}.
We show that it is possible to obtain a perfect description of 
the bandgap bowing of ternary alloys in the InGaAsSb family of materials. Furthermore, this approach is in 
a good agreement with supercell calculations using the same set of parameters.
This scheme opens a way for atomistic modeling of alloy-based opto-electronic devices 
without extensive supercell calculations.
\end{abstract}

\maketitle



The correct way to treat alloys at an atomistic level is the calculation of 
large enough supercells with randomly chosen atom distribution, and the unfolding of 
the resulting band structure, possibly averaged over realizations, to obtain an effective alloy band structure.\cite{Wei90,Boykin07,Popescu10,Rubel14,Medeiros15} 
However, this approach is hardly conceivable with \textit{ab initio} methods, and computationally expensive even in empirical-parameter methods like Empirical Tight Binding (ETB).\cite{Jancu98,Benchamekh12} Besides, for a wide range of application, it exceeds the needs of theoretical modeling: the calculation of optical and transport properties of heterostructure devices demands for a 
method which could account for atomistic features in a relatively cheap way. For modeling of simple nanostructure 
devices the computational complexity of ETB is comparable with that of effective mass approach: finite element discretization of 
$\bf{k.p}$ differential equations needs unphysically small-size elements while for atomistic approach the size of discretized Hamiltonian is of the order of the number of atoms in the structure. At the same time, ETB as an atomistic approach simplifies consideration of boundaries, accounts for atomistic interface symmetries and properties of ultra-thin layers, and allows full-band calculations.\cite{Benchamekh12}

The Virtual Crystal Approximation (VCA) which is a standard for the envelope function approaches could be an ideal trade-off between atomistic resolution and computational complexity: instead of full atomistic calculation of an alloy, a tight-binding representation 
of a virtual binary material mimicking the alloy band structure could be used in device modeling. Unfortunately, 
the connection between alloy band structure and tight-binding parametrization of the parent binary materials is not straightforward, as a significant non-linearity (or bowing) of the band parameters is observed in most material systems. This severely limits the use of ETB in device modelling. Here, we show how to construct a VCA interpolation without introducing \textit{ad hoc} parameters.

The family of materials based on InGaAsSb is a good test case for such theory: these compound materials are well documented, and widely used in modern electronic and optoelectronic devices. At the same time, they have a property which is not trivial to reproduce: band gap bowing in some of the ternary alloys is quite large, with for instance an absolute gap minimum at $x=0.4$ for the InAs$_x$Sb$_{1-x}$ alloy. In Ref.~\onlinecite{Shim98} the bowing has been accounted in tight-binding by introducing bowing of diagonal energies and transfer matrix elements. This approach gives reasonable agreement with experiments, but at the expense of introducing a number of additional parameters. 
The observed correlation between alloy bowing and lattice mismatch of parent binaries suggests that lattice mismatch is a driving effect in bowing mechanism, which we explore in the present work.


The $spds^*$ tight-binding Hamiltonian used here is constructed following Ref.~\onlinecite{Jancu98}, but with a more complete treatment of strain, in the spirit of Ref.~\onlinecite{Jancu07}. Below we describe the strain Hamiltonian for the bulk material, but this procedure may be generalized to nanostructures as explained in Ref.~\onlinecite{Raouafi15}. The strain contribution to the tight-binding Hamiltonian has three rather distinct parts. 
First contribution is a scaling of the transfer matrix elements with respect to bond length\cite{Jancu98}

\begin{equation}\label{eq:tme_strain}
  V_{n_1,n_2;ijk} = V_{n_1,n_2;ijk}^0
    \left( \frac{d_{n_1,n_2}}{d_{n_1,n_2}^0} \right)^{n_{ijk}},
\end{equation}

here $n_1$ and $n_2$ are two neighboring atoms, $ijk$ encodes corresponding Slater-Koster off-diagonal parameters, $d_{n_1,n_2}$ is the relaxed interatomic distance, $d_{n_1,n_2}^0$ is the bond length in unstrained
bulk binary compound and $n_{ijk}$ is the exponent of power-law scaling\cite{Jancu98}. Here, we use an original parametrization given in Table~\ref{tbl:strain_par}.

Another contribution, also considered in Refs.~\onlinecite{Jancu98, Jancu07}, is a shift of orbital energies proportional to hydrostatic component of strain tensor:

\begin{equation}\label{eq:hydr_strain}
E_{\beta} = E_{\beta}^0 - \alpha_{\beta} (E_{\beta}-E_{\text{ref}}) \frac{\operatorname{Tr}\varepsilon}3,
\end{equation}

where $\beta$ enumerates orbitals, $\alpha_{\beta}$  are parameters given in Table~\ref{tbl:strain_par}.
For each binary compound, we define a reference energy $E_{\text{ref}}=E_{s^*}-6E_{\left\langle1,0,0\right\rangle}$, where $E_{\left\langle1,0,0\right\rangle}=\hbar^2(2\pi/a)^2/2m_0$ and $a$ is the lattice constant. The introduction of the reference energy is particularly important as it avoids a change of the strain parameters when a band offset comes into play. The scaling to orbital energy used previously \cite{Jancu98, Jancu07} had the merit of being analytically exact in the case of a free-electron crystal where energy is purely kinetic.
The choice made for the reference energy is motivated by the aim to keep, as a first approximation, the $s^*$ orbitals the same as in the free electron limit, and thus the number of adjustable parameters is reduced by maintaining $\alpha_{s^*}=2$. $E_{\text{ref}}$ can also be understood as the average crystal potential.

 \begin{table}\caption{Strain parameters used in calculations. In addition to what is given in table, $\alpha_{s^*}=2.0$; $n_{s^*s^*\sigma}=n_{s^*d\sigma}=n_{dd\sigma}=n_{dd\pi}=n_{dd\delta}=2.0$.}
\label{tbl:strain_par}
 \begin{tabular*}{\columnwidth}{@{\extracolsep{\fill}}lrrrrr}
 \hline
 &InAs&GaAs&GaSb&InSb\\
 \hline
$   \alpha_s^a    $& $ 0.6415$& $ 0.0659$& $ 2.4155$& $ 3.9081$ \\
$   \alpha_p^a    $& $ 1.9385$& $ 1.6817$& $ 1.8949$& $ 2.0771$ \\
$   \alpha_d^a    $& $ 2.0759$& $ 2.5626$& $ 1.8885$& $ 1.7951$ \\
$   \alpha_s^c    $& $ 0.6415$& $ 0.0659$& $ 2.4155$& $ 3.9081$ \\
$   \alpha_p^c    $& $ 1.9385$& $ 1.6817$& $ 1.8949$& $ 2.0771$ \\
$   \alpha_d^c    $& $ 2.0759$& $ 2.5626$& $ 1.8885$& $ 1.7951$ \\
 \hline
$n_{ss\sigma}     $& $  4.3772$& $  5.6252$& $  2.9656$& $  4.5914$ \\
$n_{sp\sigma}     $& $  6.0795$& $  5.6059$& $  2.4114$& $  7.3228$ \\
$n_{sd\sigma}     $& $  6.7901$& $  2.4346$& $  7.6161$& $  6.4034$ \\
$n_{ss^*\sigma}   $& $  4.4281$& $  5.1500$& $  2.3726$& $  5.3162$ \\
$n_{s^*p\sigma}   $& $  4.9488$& $  6.2613$& $  6.4571$& $  3.5637$ \\
$n_{pp\sigma}     $& $  7.1545$& $  6.6889$& $  5.3818$& $  5.0427$ \\
$n_{pp\pi}        $& $  5.9041$& $  7.6765$& $  5.8046$& $  3.5717$ \\
$n_{pd\sigma}     $& $  7.6013$& $  7.7359$& $  6.1182$& $  6.6236$ \\
$n_{pd\pi}        $& $  2.0781$& $  4.7523$& $  5.8451$& $  2.8179$ \\
 \hline
$\pi_{001}        $& $  0.1746$& $  0.2704$& $  0.2206$& $  0.2175$ \\
$\pi_{111}        $& $  0.5115$& $  0.1588$& $  0.5803$& $  0.1000$ \\
 \hline
 \hline
 \end{tabular*}

 \end{table}

The third contribution is a splitting of the on-site energies of degenerate orbitals according to strain symmetry.\cite{Jancu07,Niquet09,Boykin10,Zielinski12,{Raouafi15}} 
The simplest approach for this contribution is to use a correction proportional to the strain tensor.\cite{Jancu07,Raouafi15} 
Using the method of invariants, it can be shown that the $p$-orbital same-atom block in the tight-binding Hamiltonian
(the basis functions are $p_x$, $p_y$, $p_z$) has the form:

\begin{equation}\label{ham_str_n_int}
\delta\hat{H}=
\left(\begin{array}{ccc}
\lambda_1(\sqrt3\varepsilon_1-\varepsilon_2)
   &\lambda_2\varepsilon_{xy} + \xi u_z
       &\lambda_2\varepsilon_{zx} + \xi u_y\\
\lambda_2\varepsilon_{xy}+\xi u_z
   &-\lambda_1(\sqrt3\varepsilon_1+\varepsilon_2)
       &\lambda_2\varepsilon_{yz} +\xi u_x\\
\lambda_2\varepsilon_{zx}+\xi u_y
   &\lambda_2\varepsilon_{yz}+\xi u_x
       &2\lambda_1\varepsilon_2
\end{array}\right),
\end{equation}

where we defined $\varepsilon_1 = \sqrt{3}(\varepsilon_{xx}-\varepsilon_{yy})$, $\varepsilon_2=2\varepsilon_{zz}-\varepsilon_{xx}-\varepsilon_{yy}$. Following Ref.~\onlinecite{Jancu07} we also assume $\xi=\pm\lambda_{2}$, with ``$+$'' for anion and ``$-$'' for cation. To make the parametrization space more compact, we assume that the parameters on anion and cation are connected as:

\begin{equation}
\begin{split}
  \lambda_{1\beta} & =  \frac12 (E_{p\beta}-E_{\text{ref}}) \pi_{100} \\
  \lambda_{2\beta} & = -\frac83 (E_{p\beta}-E_{\text{ref}}) \pi_{111} 
\end{split}
\end{equation}

here $\beta$ is either anion or cation and parameters $\pi_{100}$, $\pi_{111}$ are similar to parameters introduced in Ref.~\onlinecite{Jancu07}. Note that in principle, a completely similar splitting holds for the $\Gamma_{15}$ representation of d orbitals. However, for deformation potentials at $\Gamma$ the $p$- and $d$-orbitals effects can be renormalized into each other, so for simplicity, we set to zero the $d$-orbital splitting.

We further reduce the parameter space by setting coefficients $n_{ijk}$ for the nearly free-electron states $s^*$ and $d$ to the free electron value $n=2$ and by imposing regularity in the chemical dependencies.
In comparison with other strain parameterizations, our approach has relatively small number of parameters. 
Still, the determination of strain parameters for the InGaAsSb family represents a challenging task, because of the small number of available well documented deformation potentials. Indeed, the deformation potentials at the zone center do not provide enough information to allow a unique fit of the still large number of adjustable parameters. Since we expect that any parametrization that gives good values of $a_c$, $a_v$, $b$ and $d$ would provide satisfactory results for realistic device modeling, the strain parameters are numerically fitted to reproduce the recommended values of deformation
potentials in the center of Brillouin zone given in Ref.~\onlinecite{Meyer01} and we use remaining parametrical flexibility to fit the bowings, as discussed below.  
For completeness, the bulk ETB parameters are listed in Table~\ref{tbl:TB_par}. They are a slightly re-worked version of those in Ref.~\onlinecite{Jancu98}. Note that in case of a significant modification of these bulk parameters, strain parameters themselves should also be revised.

\begin{table}\caption{Bulk ETB parameters used in present calculations.} 
\label{tbl:TB_par}
 \begin{tabular*}{\columnwidth}{@{\extracolsep{\fill}}lrrrrr}
 \hline
 \hline
 &InAs&GaAs&GaSb&InSb\\
 \hline
$              a  $& $  6.0580$& $  5.6500$& $  6.0959$& $  6.4794$ \\
 \hline
$          E_{s}^a$& $ -6.4738$& $ -5.9820$& $ -6.0022$& $ -6.1516$ \\
$        E_{s^*}^a$& $ 16.8502$& $ 19.4477$& $ 16.4645$& $ 14.7582$ \\
$          E_{s}^c$& $ -0.1418$& $ -0.3803$& $ -0.6905$& $ -0.3634$ \\
$        E_{s^*}^c$& $ 16.8393$& $ 19.4548$& $ 16.4745$& $ 14.8015$ \\
$          E_{p}^a$& $  2.4784$& $  3.3087$& $  2.3961$& $  2.1150$ \\
$          E_{d}^a$& $ 11.3833$& $ 13.2015$& $ 11.1422$& $  9.8811$ \\
$          E_{p}^c$& $  5.2829$& $  6.3801$& $  5.7961$& $  5.5198$ \\
$          E_{d}^c$& $ 11.3991$& $ 13.2055$& $ 11.1469$& $  9.9511$ \\
 \hline
$         ss\sigma$& $ -1.5096$& $ -1.6874$& $ -1.4707$& $ -1.2228$ \\
$   s^*_as_c\sigma$& $ -2.0155$& $ -2.1059$& $ -1.8137$& $ -1.6619$ \\
$   s_as^*_c\sigma$& $ -1.1496$& $ -1.5212$& $ -1.2303$& $ -1.3929$ \\
$     s^*s^*\sigma$& $ -3.3608$& $ -3.7170$& $ -3.0680$& $ -2.8985$ \\
$     s_ap_c\sigma$& $  2.2807$& $  2.8846$& $  2.3357$& $  2.2046$ \\
$     s_cp_a\sigma$& $  2.6040$& $  2.8902$& $  2.5600$& $  2.3639$ \\
$   s^*_ap_c\sigma$& $  1.9930$& $  2.5294$& $  2.0586$& $  1.6962$ \\
$   s^*_cp_a\sigma$& $  2.0708$& $  2.3883$& $  2.2985$& $  1.9879$ \\
$     s_ad_c\sigma$& $ -2.8945$& $ -2.8716$& $ -2.6114$& $ -2.3737$ \\
$     s_cd_a\sigma$& $ -2.3175$& $ -2.2801$& $ -2.3460$& $ -2.1767$ \\
$   s^*_ad_c\sigma$& $ -0.6393$& $ -0.6568$& $ -0.6274$& $ -0.5548$ \\
$   s^*_cd_a\sigma$& $ -0.5949$& $ -0.6113$& $ -0.5925$& $ -0.4875$ \\
$         pp\sigma$& $  3.6327$& $  4.4048$& $  3.8153$& $  3.4603$ \\
$            pp\pi$& $ -0.9522$& $ -1.4471$& $ -1.4133$& $ -1.1630$ \\
$     p_ad_c\sigma$& $ -1.1156$& $ -1.6035$& $ -1.2955$& $ -1.3928$ \\
$     p_cd_a\sigma$& $ -1.3426$& $ -1.6260$& $ -1.8115$& $ -1.4145$ \\
$        p_ad_c\pi$& $  1.2101$& $  1.8423$& $  1.6714$& $  1.1921$ \\
$        p_cd_a\pi$& $  1.5282$& $  2.1421$& $  1.8909$& $  1.7536$ \\
$         dd\sigma$& $ -0.8381$& $ -1.0885$& $ -0.9200$& $ -0.6688$ \\
$            dd\pi$& $  1.9105$& $  2.1560$& $  1.8697$& $  1.4601$ \\
$         dd\delta$& $ -1.3348$& $ -1.8607$& $ -1.6545$& $ -1.4373$ \\
 \hline
$\Delta_a/3       $& $  0.1558$& $  0.1745$& $  0.3841$& $  0.3810$ \\
 \hline
$\Delta_c/3       $& $  0.1143$& $  0.0408$& $  0.0410$& $  0.1275$ \\
 \hline
 \hline
 \end{tabular*}

 \end{table}

We construct the tight-binding parameters of the alloy A$_x$B$_{1-x}$C from the tight-binding parameters of binary materials AC and BC using the following procedure:
First, the lattice constant of the alloy is found as a linear interpolation between binaries (Vegard's law). Then we use our description of strain (see above) to construct the parameters of materials AC and BC strained to the lattice constant of 
the alloy. Next, the tight-binding parameters of the alloy are taken as linear interpolation of the strained binary materials. 

\begin{table*}
\begin{tabular}{c|rrrrrrr}
\hline
             &  InAs & GaAs & GaSb & InSb & In$_{0.2}$Ga$_{0.8}$As & In$_{0.53}$Ga$_{0.47}$As & InAs$_{0.4}$Sb$_{0.6}$ \\
\hline
$E_g$        & $0.417$  & $1.519$ & $0.811$ & $0.234$ & $1.207$&  $0.814$& $0.1239$\\
$m_e$        & $0.0235$ & $0.066$ & $0.0402$ &$0.0132$& $0.0519$& $0.037$& $0.0069$\\
\hline                                                                    
$a_c$        & $-5.08$  & $-7.17$ & $-7.50$ & $-6.94$ & $-6.62$ & $-5.88$ & $-6.04$\\
$a_v$        & $ 1.00$  & $ 1.16$ & $ 0.80$ & $ 0.36$ & $ 1.22$ & $ 1.23$ & $ 0.40$ \\
$b$          & $-1.80$  & $-2.00$ & $-2.00$ & $-2.00$ & $-1.35$ & $-0.13$ & $-0.54$ \\
$d$          & $-4.25$  & $-5.82$ & $-5.09$ & $-5.18$ & $-5.22$ & $-5.00$ & $-4.86$ \\
\hline
\end{tabular}
\caption{Some material parameters computed using strain parameters from Table~\ref{tbl:strain_par}\label{tbl:kp_par} and ETB unstrained hamiltonian parameters from Table~\ref{tbl:TB_par}.
}
\end{table*}

\begin{figure}
\includegraphics[width=\linewidth]{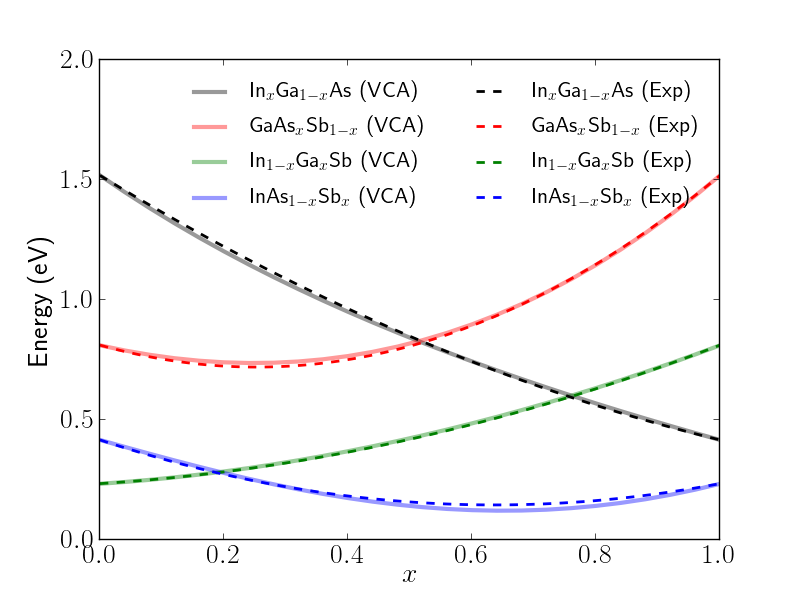}%
\caption{Band gap as a function of composition for ternary materials
calculated with the ETB interpolation (solid lines) compared 
to available experimental data (dashed lines).~\cite{Donati03}
}\label{fig:VCA}
\end{figure}

To compare with experimental data, we use the values from Ref.~\onlinecite{Donati03}. As can be seen from Fig.~\ref{fig:VCA}, the comparison is almost perfect. The  bandgap bowings are very well reproduced, and we further checked that the change of the top of valence band and bottom of conduction band separately agree with available experimental data. 
Also, we observe reasonable behavior of basic properties of alloys in VCA like effective masses and deformation potentials. For completeness, we also give in Table~\ref{tbl:kp_par} effective masses of electron and deformation potentials of the binaries and few widely used alloys.

\begin{figure}
\includegraphics[width=\linewidth]{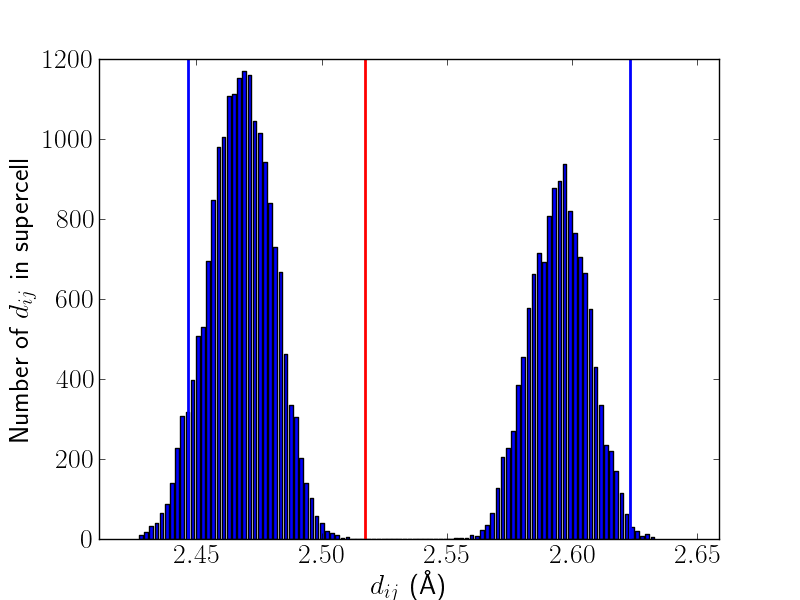}%
\caption{Distribution of bondlengths in supercell of random 
InGaAs alloy after VFF elastic energy relaxation.
Vertical blue lines show bond lengths corresponding to bulk 
InAs and GaAs, vertical red line shows the bond length of 
virtual In$_{0.4}$Ga$_{0.6}$As obtained as a linear interpolation.
}\label{fig:bonds}
\end{figure}

Despite its obvious usefulness for device modeling, the virtual crystal approximation is known to differ appreciably from reality: EXAFS  studies\cite{Mikkelsen82,Balzarotii85,Glas95} have revealed in the early 80's that in ternary alloys, individual bond lengths keep a value very close to that in corresponding unstrained binary bulk material. It is therefore of utmost interest to compare our VCA approach with supercell calculations of random alloys.
For the supercell calculation of alloys we use a $(10 a)^3$ cubic supercell containing 8000 atoms. 
We randomly distribute atoms in accordance with alloy composition, then set periodic boundary conditions fitting the alloy
lattice constant $a$ and relax atomic positions using the valence force field (VFF) approach,\cite{Keating66} which is
known to give reasonable results for small and intermediate values of strain.\cite{Steiger11}
During the relaxation we keep periodic boundary conditions fixed and change atomic positions 
using conjugate gradient method to minimize elastic energy in VFF.
After minimization, we obtain atomic positions in the fully relaxed structure. The distribution of bond lengths for a In$_{0.4}$Ga$_{0.6}$As alloy is displayed in Fig.~\ref{fig:bonds} to illustrate agreement with EXAFS measurements.~\cite{Mikkelsen82} It is easily seen that no bond in the alloy has a bond-length corresponding to the alloy lattice parameter!

\begin{figure}
\includegraphics[width=\linewidth]{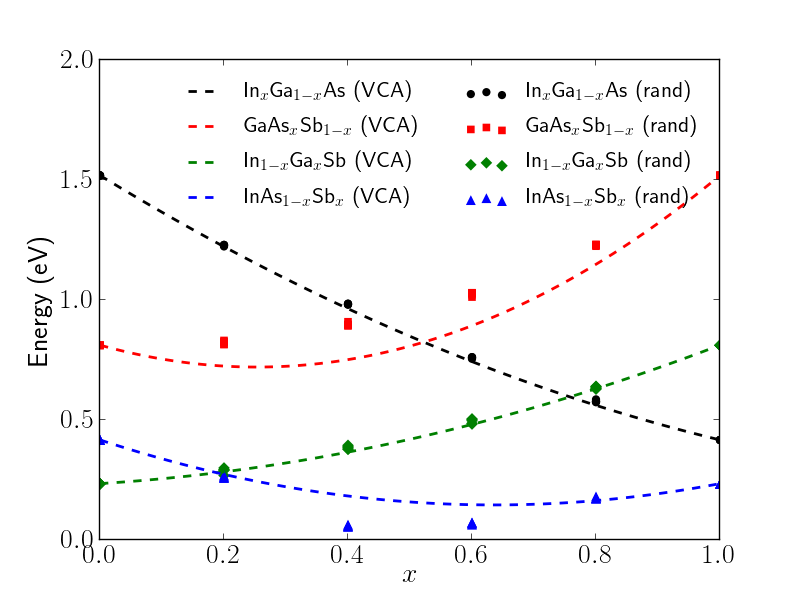}%
\caption{Band gap of random alloys (symbols)
as a function of composition for the $10 a\times 10 a\times 10 a$ supercell, compared with VCA (dashed lines).
}\label{fig:random}
\end{figure}

For this relaxed structure we extract a microscopic strain tensor using the approach explicited in Ref.~\onlinecite{Raouafi15}. 
Using relaxed atomic positions and microscopic strain tensor, we construct the matrix of ETB Hamiltonian and find few eigenvalues near the band gap using the original implementation of thick-restarted Lanczos iterations.\cite{TRLan10}
From these energies we estimate the bandgap of alloy. Its value depends on particular distribution of atoms within the considered random supercell. To estimate the error in random alloy bandgap, we compute this value for a set of five supercell realizations and compare the obtained bandgap with the VCA calculation.
The results for all four ternary alloys as a function of composition $x$ are presented in Fig.~\ref{fig:random}. Dashed lines show the VCA bandgap as a function of alloy composition and dots show the bandgap of random alloys. As can be seen, despite the fact that the latter has a complicated 
strain distribution and cannot be easily mapped to a virtual crystal, the bandgaps, computed with the same set of ETB parameters, are in close agreement. For a more detailed comparison, the unfolding procedure\cite{Wei90,Boykin07,Popescu10,Rubel14,Medeiros15} is necessary, which goes far beyond the scope of the present paper.

In conclusion, we propose a VCA description of ternary alloys from the InGaAsSb family of materials in the $spds^*$ ETB model.
The method is based on a new parametrization of strain coefficients in the binary materials and a simple yet original interpolation scheme. The present parametrization gives an accurate fit of available experimental data for the basic properties of ternary alloys, and also nicely reproduces results of random alloy simulations. Combination of VCA and tight-binding allows for the accurate modeling of nano-sized devices where atomistic details and/or full-band description is necessary.

\paragraph*{Acknowledgments.}  This work was supported by Russian-French International Laboratory ILNACS,
RFBR grant 15-32-20828 and the European Union 7th Framework Programme DEEPEN (Grant Agreement No. 604416)

\bibliography{VCA_InGaAsSb}


\end{document}